\documentclass{elsart}

\usepackage{graphics}

\usepackage{epsfig}
\usepackage{amssymb}

\newcommand{\lapproxeq}{\lower .7ex\hbox{$\;\stackrel{\textstyle
<}{\sim}\;$}}
\newcommand{\gapproxeq}{\lower .7ex\hbox{$\;\stackrel{\textstyle
>}{\sim}\;$}}
\newcommand{\stackdown}[2]{\lower 1.4ex\hbox{$\;\stackrel{\textstyle{#1}}
{\scriptstyle{#2}}\;$}}
\newcommand{\beq}{\begin{equation}}
\newcommand{\eeq}{\end{equation}}
\newcommand{\bea}{\begin{eqnarray}}
\newcommand{\eea}{\end{eqnarray}}

\def\beq{\begin{equation}}
\def\eeq{\end{equation}}

\makeatletter
\def\slash{\@ifnextchar[{\fmsl@sh}{\fmsl@sh[0mu]}}
\def\fmsl@sh[#1]#2{%
  \mathchoice
    {\@fmsl@sh\displaystyle{#1}{#2}}%
    {\@fmsl@sh\textstyle{#1}{#2}}%
    {\@fmsl@sh\scriptstyle{#1}{#2}}%
    {\@fmsl@sh\scriptscriptstyle{#1}{#2}}}
\def\@fmsl@sh#1#2#3{\m@th\ooalign{$\hfil#1\mkern#2/\hfil$\crcr$#1#3$}}
\makeatother
\def\beq{\begin{equation}}
\def\eeq{\end{equation}}

\catcode`\@=11 % This allows us to modify PLAIN macros.

\def\lsim{\mathrel{\mathpalette\@versim<}}
\def\gsim{\mathrel{\mathpalette\@versim>}}
\def\@versim#1#2{\vcenter{\offinterlineskip
    \ialign{$\m@th#1\hfil##\hfil$\crcr#2\crcr\sim\crcr } }}

\def\t1{{\tilde 1}}

\def\slash#1{#1\hskip-6pt/\hskip6pt}

\begin{document}

\begin{frontmatter}



\title{Erratum: Robust Limits on Lorentz Violation from Gamma-Ray Bursts}
\author{John~Ellis}
\address{Theory Division, Physics Department, CERN, CH-1211 Geneva 23, Switzerland}
\ead{John.Ellis@cern.ch}

\author{N.E. Mavromatos}
\address{Department of Physics, King's College London, University of London,
Strand, London WC2R 2LS, UK}
\ead{Nikolaos.Mavromatos@cern.ch}

\author{D.V.~Nanopoulos}
\address{Department of Physics, Texas A \& M University, College Station,
TX~77843, USA \\ Astroparticle Physics Group, Houston Advanced Research Center (HARC),
Mitchell Campus, Woodlands, TX~77381, USA \\ Academy of Athens,
Division of Natural Sciences, 28~Panepistimiou Avenue, Athens 10679,
Greece}
\ead{Dimitri.Nanopoulos@cern.ch}

\author{A.S. Sakharov\corauthref{cor}}
\address{Theory Division, Physics Department, CERN, CH-1211 Geneva 23, Switzerland \\
Swiss Institute of Technology, ETH-Z\"urich, 8093 Z\"urich,
Switzerland}
\corauth[cor]{Corresponding author.}
\ead{Alexandre.Sakharov@cern.ch}

\author{E.K.G.~Sarkisyan}
\address{EP Division, Physics Department, CERN, CH-1211 Geneva 23, Switzerland \\
Physics Department, Universiteit Antwerpen, B-2610 Wilrijk, Belgium}
\ead{Edward.Sarkisyan@cern.ch}

\begin{abstract}
We correct the fitting formula used in refs.~\cite{grb,grbaa} to obtain a robust limit on a violation of Lorentz invariance that depends linearly on the photon energy. The correction
leads to a slight increase of the limit on the scale of the violation, to  $M > 1.4 \times 10^{16}$~GeV.
\end{abstract}


\end{frontmatter}

It has been resently pointed out in~\cite{piran} that, due to the fact that
the comoving distance that light travels while coming from
an object at redshift $z$ in the expanding Universe is bigger by a factor $(1+z)$
than the proper distance~\cite{piran1}, formula (1) in~\cite{grb} (see
also formula (13) in~\cite{grbaa}) for the
difference in the
arrival times of two photons with energies differing by  $\Delta E$ in the case of a
linear violation of Lorentz invariance should be corrected to read:
\beq
\label{timedel1}
\Delta t_{\rm LV}=H_0^{-1}\frac{\Delta E}{M}\int\limits_0^z\frac{(1+z)dz}{h(z)},
\eeq
where $H_0$ is the Hubble expansion rate,
\beq
\label{h}
h(z) = \sqrt{\Omega_{\Lambda} + \Omega_M (1 + z)^3},
\eeq
and we assume a spatially-flat
Universe: $\Omega_{\rm total} = \Omega_{\Lambda} + \Omega_M = 1$ with
cosmological constant $\Omega_{\Lambda} \simeq 0.7$.

\begin{figure}[t]
%\centerline{\psfig{file=fit_rescaled_bw.eps,height=7cm,width=16cm}}
\centerline{\psfig{file=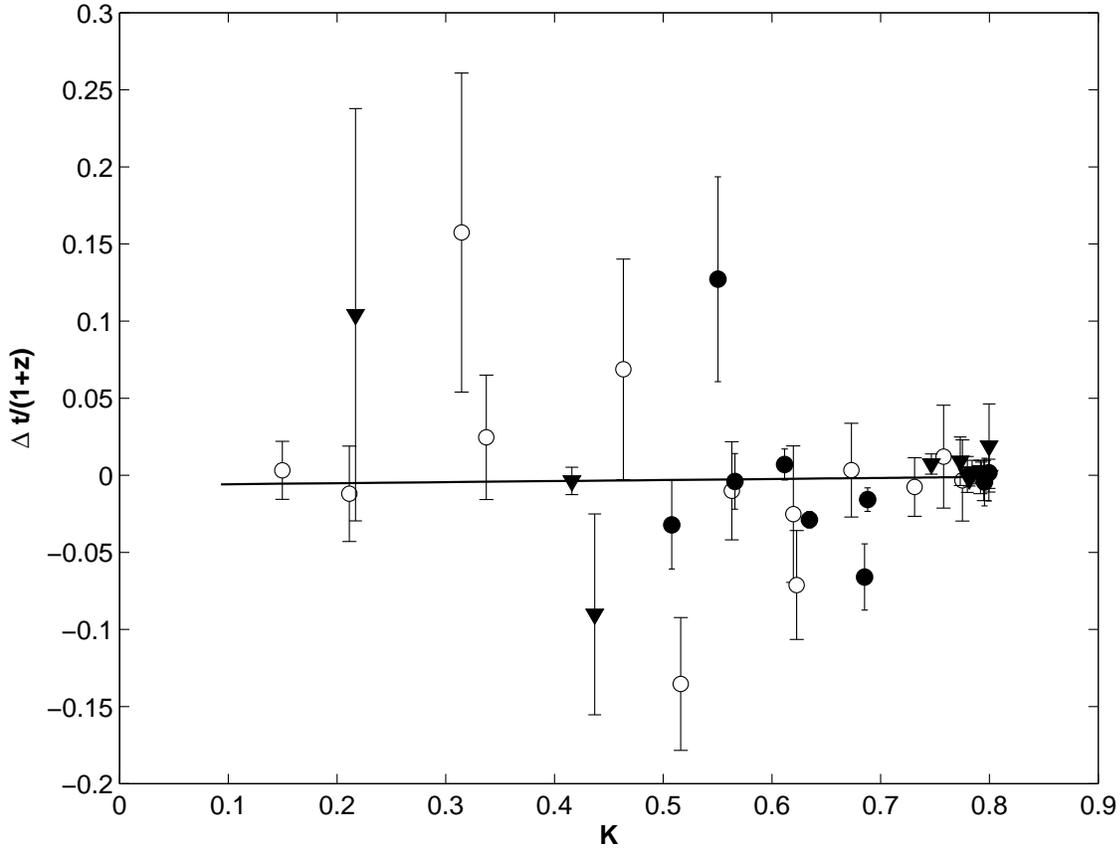,width=15cm}}
\vspace*{8pt}
\caption{
{\it
The rescaled
spectral time-lags between the arrival times of pairs of genuine high-intensity
sharp features detected in the light curves of the full set of 35 GRBs with
measured redshifts observed by BATSE (closed circles), HETE (open circles)
and SWIFT (triangles).}}
\label{regr}
\end{figure}

\begin{figure}[t]
\centerline{\psfig{file=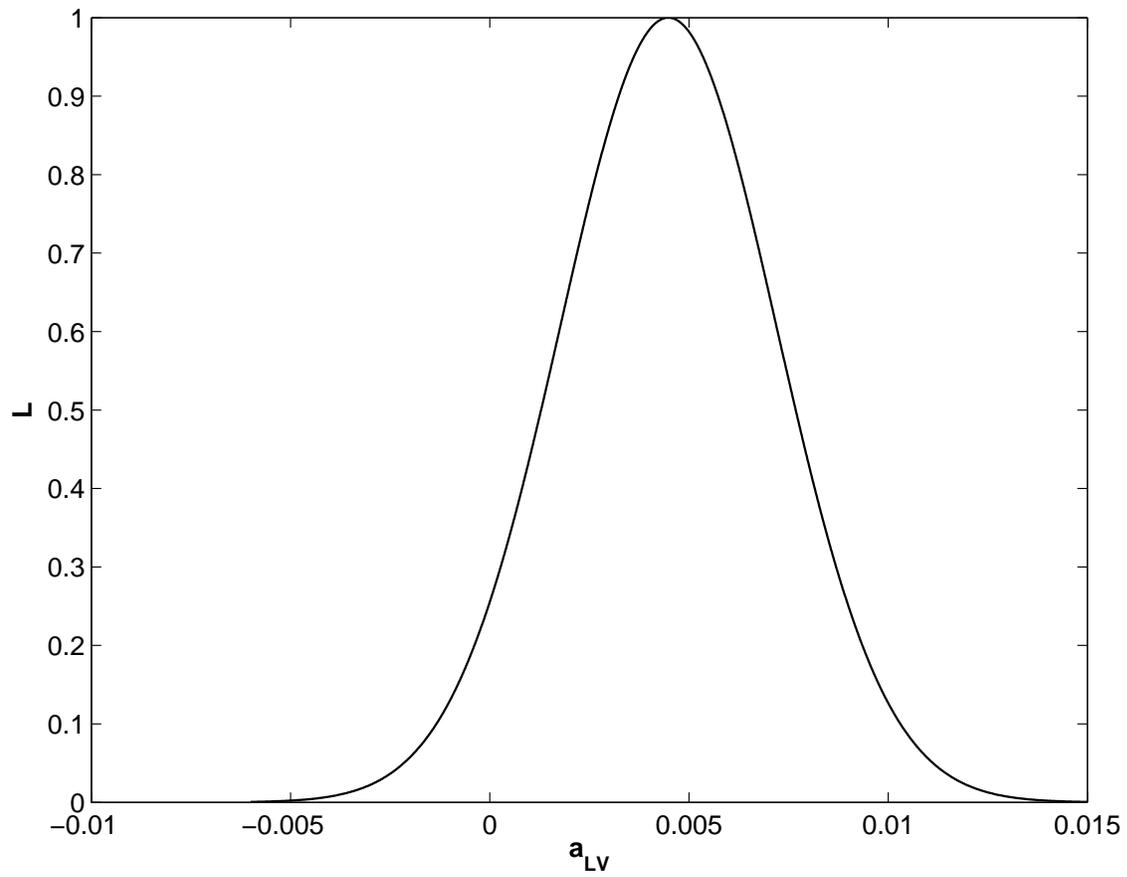,width=15cm}}
\vspace*{8pt}
\caption{
{\it
The likelihood function for the slope parameter.}}
\label{lkl}
\end{figure}

As a result of this correction, the arrival time delays calculated in~\cite{grb}
should be fitted by a linear function, as in equation (4) of~\cite{grb}
but in terms of the variable:
\beq
\label{K1}
K \equiv \frac{1}{1+z}\int\limits_0^z\frac{(1+z)dz}{h(z)}.
\eeq
The fit replacing the left panel of Fig.~2 in~\cite{grb} is presented in
Fig.~\ref{regr}. The linear fit corresponds to
\beq
\label{fit_reduced}
\frac{\Delta t_{\rm obs}^{\rm dw}}{1+z} \;
= \; (0.0068 \pm 0.0067) \, K\; - \; (0.0065 \pm 0.0046),
\eeq
and the likelihood function for the slope parameter analyzed in equation (14) of~\cite{grb} is
presented in Fig.~\ref{lkl} and, in fact, reflects better the sensitivity of
the fit and in this sense replaces Fig.~4 of~\cite{grb}.

%\begin{figure}[t]
%\centerline{\psfig{file=fit_direct_bw.eps,width=16cm}}
%\vspace*{8pt}
%\caption{
%{\it The direct fit.}}
%\label{dir}
%\end{figure}

The 95\% confidence-level lower limit obtained by solving equation
(14) of~\cite{grb} is
\beq
\label{linlimit}
M\ge 1.4 \times 10^{16}~{\rm GeV},
\eeq
compared with our previous limit $M \ge 0.9 \times 10^{16}$~GeV.

%\bigskip
\bigskip
We thank Tsvi Piran for his communication on the subject.

\end{document}